\documentclass[epjST,draft]{svjour}
\usepackage{graphicx}
\usepackage{epsfig}
\usepackage{mathtools}
\usepackage{color}
\usepackage{sistyle}
\usepackage{subfig}
\usepackage{calc}

\title{Comparison of Different Parallel Implementaions of the 2+1-Dimensional 
KPZ Model and the 3-Dimensional KMC Model}
\author{Jeffrey~Kelling\inst{1} \and G\'{e}za~\'{O}dor\inst{2} 
\and M\'{a}t\'{e} Ferenc~Nagy\inst{3} \and Henrik~Schulz\inst{4} \and
Karl-Heinz~Heinig\inst{1}}

\institute{Helmholtz-Zentrum Dresden-Rossendorf, 
Institute of Ion Beam Physics and Materials Research, 
P.O.Box 51 01 19, 01314 Dresden, Germany
\and Research Centre for Natural Sciences, Hungarian Academy of Sciences  
MTA TTK MFA, P. O. Box 49, H-1525 Budapest, Hungary
\and Wigner Research Centre for Physics, Hungarian Academy of Sciences, 
P.O.Box 49, H-1525 Budapest, Hungary
\and Helmholtz-Zentrum Dresden-Rossendorf, Department of Information Technology, 
P.O.Box 51 01 19, 01314 Dresden, Germany}

\abstract{We show that efficient simulations of the Kardar-Parisi-Zhang 
interface growth in $2+1$ dimensions and of the $3$-dimensional 
Kinetic Monte Carlo of thermally activated diffusion can be realized 
both on GPUs and modern CPUs. 
In this article we present results of different implementations 
on GPUs using CUDA and OpenCL and also on CPUs using OpenCL and MPI. 
We investigate the runtime and scaling behavior on different
architectures to find optimal solutions for solving current simulation 
problems in the field of statistical physics and materials science.}

\begin{document}
\input epsf.sty
\maketitle

\section{Introduction}

Statistical physics and materials science use advanced simulation tools 
to understand complex phenomena prevalent in nature.
To analyze the behavior in the thermodynamic limit we need to reach 
extremely large system sizes and times.
Simulations of disordered systems require several hundreds of hours of 
computing time due to the slow evolution even in one dimension \cite{cpc11}.

In present day parallel computing architectures the efficiency of the 
parallelization is in the focus of software development. 
Current approaches to write parallel algorithms can be divided into 
two groups: on one hand thread-parallel algorithms on GPUs 
(CUDA or OpenCL) and CPUs (OpenCL or OpenMP) 
and messages-based process-parallel algorithms on the other hand. 
In applications both concepts can be combined, since threads can only be 
created on the motherboard, whereas the communication between different 
units \footnote{different can also mean architectural inhomogeneities} 
can only be realized using message passing.

In this article we investigate two different models,  
the Kardar-Parisi-Zhang (KPZ) surface growth and the Kinetic Monte Carlo (KMC) 
of thermally activated diffusion of binary alloys, implemented using different 
programming models on different architectures. 
Main specifications of the GPUs used in this work are gathered in 
Tables \ref{tableGPU1} and \ref{tableGPU2}, while the most important details of the CPUs
utilized for comparison can be found in Table \ref{tableCPU}. 
Note, that the performance values in both tables are theoretic.

\stepcounter{footnote}
\footnotetext{64 kB can be split into 48 kB of shared memory and 16 kB of L1
cache or vice versa.}
\begin{table}
\caption{Overview of the key facts of the NVIDIA GPUs used.\label{tableGPU1}}
\begin{center}
\begin{tabular}{lrr}
\hline
& NVIDIA C1060 & NVIDIA C2050 / C2070 \\
& (Tesla) & (Fermi) \\
\hline
Number of multiprocessors (mp) & 30 & 14 \\
Number of processing elements  & 240 & 448 \\
Clock rate of the mp & 1300 MHz & 1150 MHz \\
Global memory & 4 GB & 2 GB / 6 GB \\
Shared memory per mp & 16 kB & 48 kB${}^{\thefootnote}$ \\
Memory clock rate & 800 MHz & 1500 MHz \\
Global memory bandwidth & 102 GB/s & 144 GB/s \\
Peak performance (single precision) & 936 GFlop/s & 1030 GFlop/s \\
Peak performance (double precision) & 78 GFlop/s & 515 GFlop/s \\
\hline
\end{tabular}
\end{center}
\end{table}

\begin{table}
\caption{Overview of the key facts of the ATI GPUs used.\label{tableGPU2}}
\begin{center}
\begin{tabular}{lrr}
\hline
& ATI Radeon HD5970 & ATI Radeon HD6970 \\
\hline
Number of multiprocessors (mp) & 40 & 24 \\
Number of processing elements  & 3200 & 1536 \\
Clock rate of the mp & 725 MHz & 880 MHz \\
Global memory & 2 GB & 2 GB \\
Shared memory per mp & 32 kB & 32 kB \\
Memory clock rate & 1000 MHz & 1375 MHz \\
Global memory bandwidth & 256 GB/s & 256 GB/s \\
Peak performance (single precision) & 4640 GFlop/s & 2703 GFlop/s \\
Peak performance (double precision) & 928 GFlop/s & 675 GFlop/s \\
\hline
\end{tabular}
\end{center}
\end{table}

\begin{table}
\caption{Overview of the key facts of the CPUs used.\label{tableCPU}}
\begin{center}
\begin{tabular}{lrrr}
\hline
& AMD Opteron & Intel Core i5 & Intel Core i7 \\
& F8380 & 430 M & 920 \\
\hline
Number of cores 	& 4 			& 2 			& 4 \\
Clock rate 		& 2500 MHz 		& 2267 MHz 		& 2664 MHz \\
L1 cache 		& 4 $\times$ 128 kB 	& 2 $\times$ 64 kB 	& 4 $\times$ 64 kB \\
L2 cache 		& 4 $\times$ 512 kB 	& 2 $\times$ 256 kB	& 4 $\times$ 256 kB \\
L3 cache 		& 6 MB 			& 3 MB			& 8 MB \\
Peak performance 	& 49,16 GFlop/s 	& 39,68 GFlop/s		& 88,97 GFlop/s \\
\hline
\end{tabular}
\end{center}
\end{table}

In Section 2 we present the two models used in this work: the KPZ growth
and the KMC method of binary alloys. Details about the implementation of 
these two models are presented in Section 3. In Section 4 we show the efficiency 
results of the different implementations and we conclude with Section 5.

\section{Overview of the models}
\subsection{The Kardar-Parisi-Zhang model} 

The Kardar-Parisi-Zhang (KPZ) equation was inspired in part by the the 
stochastic Burgers equation \cite{Burgers74}, which belongs to the same 
universality class \cite{forster77} and it became the subject of many 
theoretical studies \cite{HZ95,barabasi,krug-rev}. 
Besides, it models other important physical phenomena such as directed 
polymers \cite{kardar85}, randomly stirred fluid \cite{forster77}, 
dissipative transport \cite{beijeren85,janssen86} and the magnetic 
flux lines in superconductors \cite{hwa92}.
Due to the mapping onto the Asymmetric Exclusion Process (ASEP)
\cite{Rost81} it is also a fundamental model of a
non-equilibrium particle system \cite{Obook08},
with broken detailed balance condition
\begin{equation}\label{dbal}
 P(C) R_{C\to C'} \ne P(C') R_{C'\to C}
\end{equation}
where $P(C)$ denotes the probability of the state $C$ and $R_{C\to C'}$
is the transition rate between states $C$ and $C'$.

The KPZ equation specifies the evolution of the height function
$h(\mathbf{x},t)$ in the $d$ dimensional space
\begin{equation}  \label{KPZ-e}
\partial_t h(\mathbf{x},t) = v + \sigma\nabla^2 h(\mathbf{x},t) +
\lambda(\nabla h(\mathbf{x},t))^2 + \eta(\mathbf{x},t) \ .
\end{equation}
Here $v$ and $\lambda$ are the amplitudes of the mean and local growth
velocity, $\sigma$ is a smoothing surface tension coefficient and $\eta$
roughens the surface by a zero-average, Gaussian noise field exhibiting
the variance
\begin{equation}  \label{Gnoise}
\langle\eta(\mathbf{x},t)\eta(\mathbf{x^{\prime}},t^{\prime})\rangle = 2 D
\delta^d (\mathbf{x-x^{\prime}})(t-t^{\prime}) \ .
\end{equation}
The letter $D$ denotes the noise amplitude and $\langle\rangle$
means distribution average.
The equation is solvable in $\left( 1+1\right) d$ due to the Galilean
symmetry \footnote{The invariance of Eq. (\ref{KPZ-e}) under an infinitesimal
tilting of the interface}, \cite{forster77} and an incidental
fluctuation-dissipation symmetry \cite{kardar87}, while in higher
dimensions approximations are available only. The model exhibits diverging
correlation length, hence a scale invariance, 
that can be understood by the particle current in the ASEP model. 
The current corresponds to the up-down anisotropy of the KPZ. 
Therefore KPZ equation has been investigated by
renormalization techniques \cite{SE92,FT94,L95}.
The KPZ phase space has been the subject of controversies for a long 
time \cite{MPPR02,F05} and the strong coupling fixed point has been 
located by non-perturbative RG very recently \cite{CCDW11}.
Values of the surface scaling exponents for $d>1$ exhibit considerable 
uncertainties (see \cite{barabasi}), we provided very high precision 
simulation results in \cite{asepddcikk,GPU2cikk}. 

Discretized versions of KPZ have also been studied a lot
(\cite{MPP,MPPR02,Reis05}, for a review see \cite{barabasi}).
Recently we have shown \cite{asep2dcikk,asepddcikk} that the 
mapping between a restricted solid on solid representation of
the KPZ surface growth and the ASEP \cite{kpz-asepmap,meakin} can
straightforwardly be extended to higher dimensions.
In 2+1 dimensions the mapping is just the simple extension of the
rooftop model to the octahedron model as can be seen on Figure~2 of
\cite{asep2dcikk}.
The surface built up from the octahedra can be described by the
edges meeting in the up/down middle vertexes. Up edges in the
$x$ or $y$ directions are approximated by the 
derivatives $\sigma_{x/y} = +1$, while the down ones by $\sigma_{x/y}= -1$.
Note, that in a renormalizable system, such as the KPZ different
slopes without overhangs can approximated on this way.
This can also be understood as a special $2d$ cellular automaton
\cite{Wolfram} with the generalized Kawasaki updating rules
\begin{equation}\label{rule}
\left(
\begin{array}{cc}
   -1 & 1 \\
   -1 & 1 
\end{array}
\right)
\overset{p}{\underset{q}{\rightleftharpoons }}
\left(
\begin{array}{cc}
   1 & -1 \\
   1 & -1 
\end{array}
\right)
\end{equation}
with probability $p$ for attachment and probability $q$ for detachment.
We have confirmed that this mapping, using the parametrization:
$\lambda = 2 p/(p+q)-1$, reproduces the one-point functions of the
continuum model.
This kind of generalization of the ASEP model can be regarded
as the simplest candidate for studying KPZ in $d>1$: a
one-dimensional model of self-reconstructing $d$-mers
\cite{BGS07} diffusing in the $d$-dimensional space.
Furthermore this lattice gas can be studied by very efficient
simulation methods.

\begin{figure}[ht]
\begin{center}
\epsfxsize=100mm
\epsffile{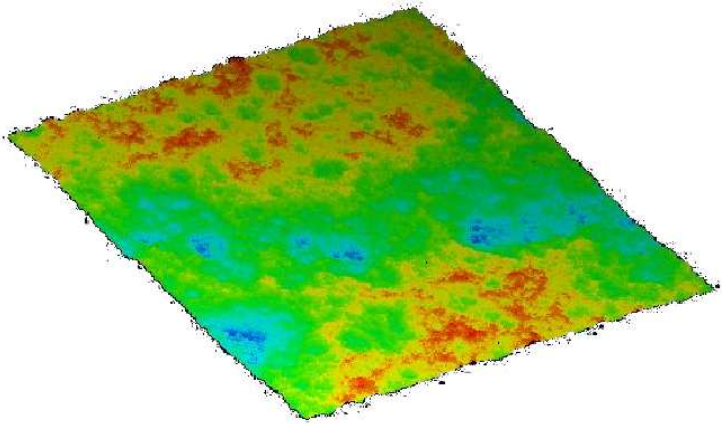}
\end{center}
\caption{Snapshot of the simulated KPZ surface using color codes.
\label{fig:surf}}
\end{figure}

We followed the evolution of the lattice gases
of linear size ($L$), started from flat initial configuration.
Periodic boundary conditions are applied.  
The surface heights (see Fig.~\ref{fig:surf})
are reconstructed from the slopes
\begin{equation}
h_{i,j} = \sum_{l=1}^i \sigma_x(l,1) + \sum_{k=1}^j \sigma_y(i,k) \ \ ,
\end{equation}
and the squared interface width
\begin{equation} 
\label{Wdef}
W^2(L,t) = \frac{1}{L^2} \, \sum_{i,j}^L \,h^2_{i,j}(t)  -
\Bigl(\frac{1}{L} \, \sum_{i,j}^L \,h_{i,j}(t) \Bigr)^2 \ .
\end{equation}
was calculated at certain sampling times ($t$).
The $W^2(L,t)$ results are written out during the run
to the disk and analyzed later by statistical methods
as discussed in \cite{GPU2cikk}.

\subsection{Kinetic Monte Carlo\label{s:3dkmc}} 

The Kinetic Monte Carlo (KMC) Method models atomistic on-lattice
dynamics on large spatio-temporal scales~\cite{strobel1999}. Commonly atomistic simulations are
performed by solving Hamilton equations for a system (i.\,e.~Molecular Dynamics).
The general idea of KMC is to use a thermodynamic model to average out 
microscopic fluctuations, creating a probabilistic model of 
on-lattice particle movement.
This model has been successfully applied to a variety of phenomena of
self-organization, for Ostwald ripening, investigations of the
Plateau Rayleigh instability~\cite{roentzsch2007} and phenomena in systems driven
by ion bombardment, like the creation of surface ripples~\cite{liedke2011} and
inverse Ostwald ripening~\cite{heinig2003}. Our GPU implementation puts
simulations at experimental spatio-temporal scales within reach.

Kinetic Lattice Monte Carlo employs a totalistic stochastic probabilistic cellular
automaton~\cite{wolframNew}, in the present case based on the nearest neighbor Ising model with Kawasaki
dynamics~\cite{kawasaki1966}. System evolution based, for example, on the interatomic many-body RGL
potential~\cite{rosato1989} can be treated too. In this work we will focus on the case of a binary
alloy containing two species A and B, encoded as single bits $0$ and $1$,
respectively. To make the model valid for most metals and to get a good approximation
for amorphous materials a face centered cubic simulation lattice is
used~\cite{mueller05}, where
each particle has twelve nearest neighbors. The simulation lattice is stored as
a sub-lattice of a simple cubic lattice~\cite{strobel1999}, where valid \texttt{fcc} coordinates
are identified by
\begin{equation}
 0 = (x\oplus y\oplus z)\wedge 1 \quad,
 \label{eq:fcc}
\end{equation}
where $\oplus$ denotes the logic bit-wise XOR.

The cellular automaton follows the Metropolis algorithm~\cite{Met}, where
species B is regarded active, while species A provides a surrounding matrix which is
passive. The role of species A and B can easily be exchanged by a particle--hole
transformation of the Hamiltonian. Through the course of the simulations $N$ update attempts are
called one Monte Carlo Step (MCS) in a lattice containing $N$ sites. 
The simulation time is measured in this unit,
which only gains physical meaning for large times~\cite{strobel1999}.

In an update attempt~\cite{roentzsch2007} a random lattice site $i$ is chosen. If
the chosen initial site $i$ is not occupied by a specimen of B the attempt is
finished, otherwise a random nearest neighbor is chosen as the final site $f$.
If site $f$ is occupied by the other species the content of sites is exchanged 
according to the Metropolis transition probability
\begin{equation}
 W_{if} = \begin{cases}
  \Gamma_{if} & n_f \geq n_i \\
  \Gamma_{if}\exp\left[ - (n_i - n_f)\cdot \varepsilon \right] & n_f < n_i \ \ ,
 \end{cases} \label{eq:metropolis}
\end{equation}
where $n_i$ and $n_f$ are the numbers of nearest neighbor sites of $i$ and $f$,
respectively, occupied with atoms of species B, $\Gamma_{if}$ is an effective jump frequency,
incorporating an activation energy barrier for the transition and 
$\varepsilon$ is the effective temperature. In our present work we set
\begin{equation}
\Gamma_{if}\equiv 1 .
\end{equation}

\section{Implementation of the Models}

When we parallelize a stochastic cellular automaton algorithm, 
to which both models discussed here resemble, 
the basic idea is to find a way of performing multiple updates independently. 
The main task is to generate a Markov chain of states, requiring site 
updates to be statistically independent.
This can be achieved by domain decomposition: the system is divided
into sets of non-interacting domains. Each domain of a set is assigned to 
a worker temporarily, while other sets remain inactive. 
The simplest decomposition scheme is the checkerboard 
decomposition~\cite{preis09}, but we have chosen different methods.

\begin{figure}
 \centering
 \subfloat[dead border decomposition]{
 \epsfxsize=60mm
 \epsffile{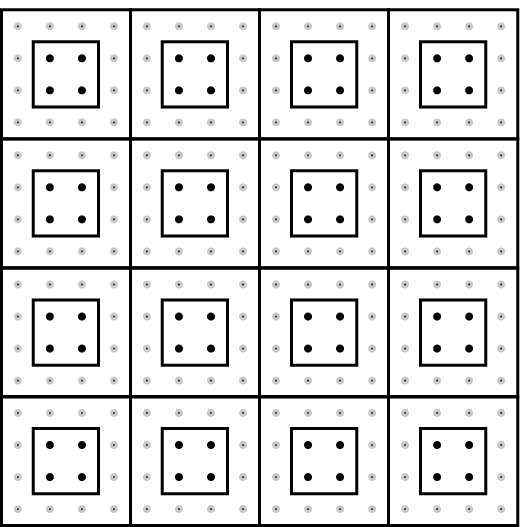}
 \label{fig:db}
 }
 \subfloat[double tiling]{
 \epsfxsize=60mm
 \epsffile{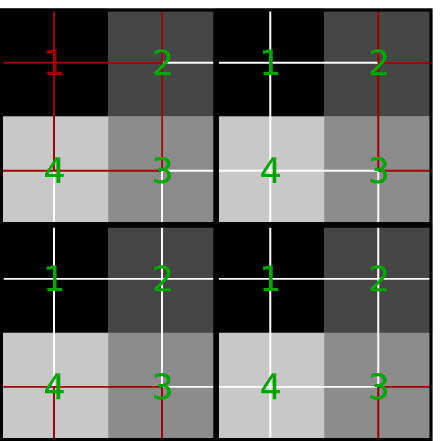}
 \label{fig:dt}
 }
 \caption{\protect\subref{fig:db}~Schematics of the dead-border decomposition. 
  Black dots denote active sites, gray dots correspond to sites 
  left out of update in a given cell decomposition.
  \protect\subref{fig:dt}~Two dimensional representation of the double tiling decomposition
  scheme. All workers update the cell of one set $\in [1,4]$ at a time. The red
  lines indicate the area that can be accessed by the worker updating the top
  left cell of set $1$ without conflicting with other workers.
 }
\end{figure}

Dead border decomposition is a scheme that has already been successfully 
employed for KPZ~\cite{GPU2cikk}. The system is decomposed into blocks,
updated independently leaving out their border. After some time the
origin of the decomposition is moved randomly to allow changes 
in the individual cells to propagate through the whole lattice
(see Fig.~\ref{fig:db}).

Another scheme, more suitable when well aligned memory accesses are
important, is the \emph{double tiling} method. The system is decomposed into
tiles, bisected in each direction, creating $2^d$ sets of independent domains.
These sets are updated in turn, generally by a randomized sequence, 
and each domain of the currently active set is assigned to a 
different worker (Fig.~\ref{fig:dt}).
This approach was also used in~\cite{shimAmar05} for a two dimensional multi-CPU
implementation of different variation of KMC.

By employing domain decomposition one deviates from the original model. 
This leads to errors at domain boundaries, which cannot be eliminated
completely, but one must keep them sufficiently small.
When a cell is updated it temporarily becomes a separate system with fixed
boundary conditions determined by the neighbors.
For sufficiently small times, this is a good approximation to a part of
a system continuously interacting with the surroundings. 

We used domain decomposition at each layer of the parallelization
independently.
On a single GPU there are two layers to be taken into account.
The \emph{device layer}, where the system has to distributed over the compute
units (work-groups in OpenCL terminology, thread block in CUDA) and the
\emph{work-group layer}, where the cell assigned to a work-group is distributed
among the threads (or work-items in OpenCL). See~\cite{weigel11} for a
overview of GPU architecture.

\subsection{The $2+1$ dimensional KPZ algorithm} 

We implemented the $2+1$ dimensional KPZ both using CUDA and OpenCL.
Our analysis was restricted to the $p=1$, $q=0$ case, while the
code could easily handle more general conditions.
An earlier version of our CUDA implementation was presented
in~\cite{GPU2cikk}, where dead border decomposition was applied at both
layers. Here we improve that application by employing a more 
efficient, single--hit double tiling scheme at the work-group layer, 
while the device layer remains unchanged. 
Because the problem is two dimensional there are $2^2$ sets of, quite small,
non-interacting cells. 
Performing full updates here,
i.\,e.~giving all sites of the cell the chance to be updated once before moving
on the next cell, would allow the effects of fixing neighboring cells to become
significant. Under these conditions the aforementioned approximation of
temporarily treating the cell as a system with fixed boundaries would become
bad.
Single--hit means, that a cell receives only a single update attempt before
the work-group, i.\,e.~the threads collectively, move on to another random set
of cells (it may be the same set). 
Single--hits repeated until, on average, each site of the work-group's 
block had the chance to be updated once. Performing single--hit updates 
effectively eliminates errors at domain boundaries, only leaving 
simultaneous updates slightly correlated.

This implementation was straightforwardly ported to OpenCL, showing almost no
difference in performance on NVIDA Tesla C2070 cards. It is however optimized
for NVIDA's architectures and thus cannot make optimal use of AMD devices. The
main difference between the two architectures, 
connected to our applications is that NVIDIA
provides 32--bit scalar registers, while AMD uses 128--bit registers
build for vector operations. AMD devices can only be fully utilized by issuing
operations on vectors of four 32--bit values. AMD devices allow different
instructions for different vector components, a technology called very long
instruction word (VLIW). The compiler may use this by automatically vectorizing
code that contains independent operations. Our CUDA implementation does not use
vector operations and leaves no room for auto vectorization, thus it can
only utilize such a device to a quarter.

\begin{figure}[ht]
\begin{center}
\epsfxsize=60mm
\epsffile{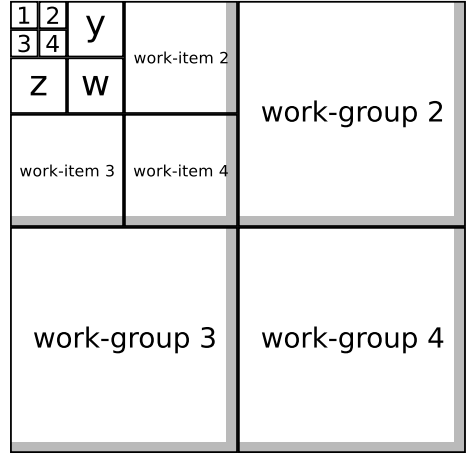}
\end{center}
\caption{Decomposition of the whole system into work-groups at device layer,
with gray areas indicating dead borders. Further decomposition at work-group
layer using double tiling is indicated for work-group $1$: Each work-item
executes four virtual threads using VLIW vector operations. The virtual threads
are denoted by their corresponding vector components ($x,y,z,w$). A single--hit
double tiling scheme is employed to distribute the work-group among all virtual
threads. The cells of the four sets of domains are indicated for virtual thread
$1.x$.
\label{fig:dc}}
\end{figure}

We created an OpenCL implementation optimized for AMD devices vectorizing the
code by hand. The basic approach was to utilize the vector capabilities of the
device by executing a \emph{virtual} thread in each vector component. At device
layer dead border decomposition is employed. At work-group layer a worker is
identical to a virtual thread. There double tiling is used to distribute the
work-group's chunk among all virtual threads. (Fig.~\ref{fig:dc})

As in the CUDA implementation the work-group layer updates are single--hit.
The difference is, that each work-item is assigned four
domains of the collectively chosen set. These four update are then carried out
using vector operations, thus achieving a maximum utilization of the ALU.

For random number generation we used different algorithms:
32-bit linear congruential (LCRNG), skip-ahead 64-bit LCRNG 
\cite{weigel11} and Mersenne Twister \cite{MT}. Comparing them by
very extensive KPZ simulations (several weeks of test runs) have shown
no noticeable differences in the scaling results \cite{GPU2cikk}.
Our OpenCL implementation employs a special version of the Mersenne Twister
called TinyMT~\cite{tinymt} for random number generation.

\subsection{Implementation of KMC} 

The GPU implementation of KMC employs a two-layer double tiling domain
decomposition scheme tailored to the two-layered computing architecture
of GPUs. At device layer the system is tiled, each tile consisting of eight blocks ($2^3$). 
Subsets of the blocks are fully updated in a sequence randomized at each MCS,
where each block of the current set is assigned to a work-group. 
For performance reasons the system size is restricted to
powers of two, leading to a number of blocks which itself is a power of
two. 
However, the number of multiprocessors on available GPUs is not a power of two.
To compensate for that (to maximize utilization), part of the super blocks are
updated ahead of time.\footnote{This does not introduce an error since the unit
MCS only has physical meaning in the limit of large times, where time ordering
is broken at the scale of two MCS.} 
At work-group layer the same decomposition scheme
as of the optimized KPZ implementation is
used:\footnote{Actually KPZ inherited this scheme from KMC.} 
double checkerboard with single hit updates.

On the GPU all threads of a block have to be synchronized, 
so there is no benefit from earlier termination of a thread, 
this would just leave part of the device idle. 
Since this happens frequently if only one species is considered to be active
(see Section \ref{s:3dkmc}), both species are considered to be
active in the GPU code.
Equation \eqref{eq:metropolis} can still be used for this, 
only the roles of the initial and final sites are reversed if 
$i$ is occupied by A.
The CUDA implementation was directly ported to OpenCL, as for the KPZ CUDA
implementation almost no difference in performance has been found on a C2070.

The condition of detailed balance~\cite{newman99} is satisfied locally up to the
work-group layer. When stepping through the domain sets at device level,
detailed balance is broken for the last few updates performed within the
individual cells, because they cannot be reversed instantly. In any way this
effect is too small to give rise any measurable effect, small disturbances of
kinetics directly at domain boundaries are of far greater
concern~\cite{kelling2012}.

The CPU MPI implementation of KMC uses dead border decomposition scheme in one
dimension. This reduces the communication overhead by improving the surface 
to bulk ratio of cells, which limits the communication of each node to its 
neighbors. The downside of this method is a lower number of workers
that can be used for fixed system size as compared to a method 
with decomposition along more dimensions. 
There is a lower limit for lateral chunk sizes one
cannot go below without hurting the statistics.

Benchmarks comparing GPU and CPU implementations exposed a problem with the CPU
implementations when dealing with very large systems for both KMC and KPZ. 
Since sites to be updated are chosen randomly the CPU can 
only make use of it's caches as long as the whole system fits at 
least in the last level cache.\footnote{For
Intel CPU's last level means L3, for AMD L2.} If the system size exceeds the
cache size the cache becomes effectively useless, causing 
a significant drop in the CPU performance.~\cite{GPU2cikk}

This can be avoided by decomposing the system into blocks, which are
updated randomly. The largest performance gain can be achieved
when those blocks are not larger that half of the size of the L1 cache. 
This can also be done in an MPI implementation. The benefit is less in smaller
systems, respectively smaller domains, for an already parallel implementation.

\begin{figure}[h!t]
\begin{center}
\epsfxsize=120mm
\epsffile{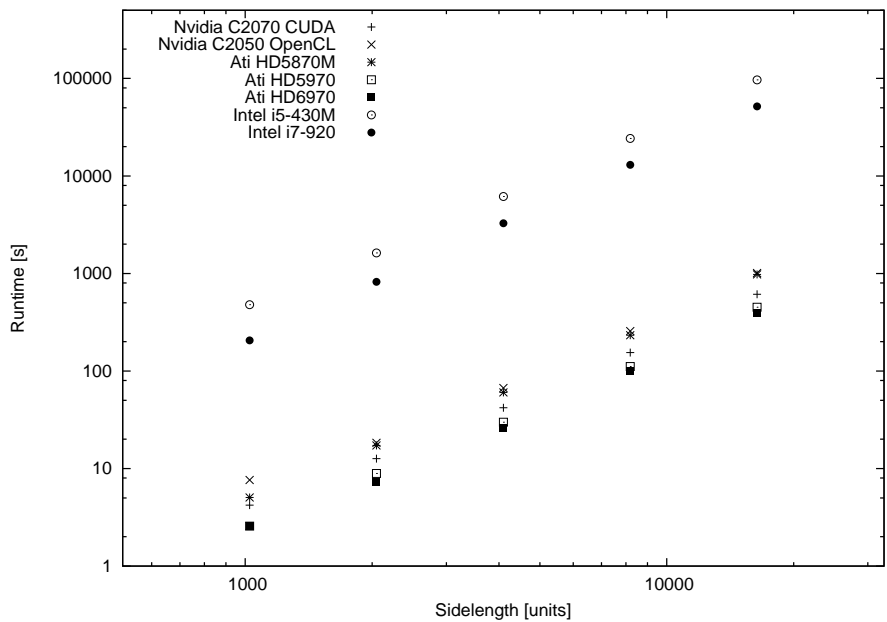}
\end{center}
\caption{Run-time comparison of the different architectures 
and programming models. For the KPZ model. The Run-time scales with the lateral
system dimension as \mbox{$\sim L^{\num{1.855}}$}.
\label{fig:plot1}}
\end{figure}

\section{Run-time comparison} 

We implemented different versions of the KPZ and KMC models 
using CUDA, OpenCL and MPI for the different platforms 
mentioned in Tables~\ref{tableGPU1}, \ref{tableGPU2} and~\ref{tableCPU}. 
In Figure~\ref{fig:plot1} we collected the run-times of a given 
task measured on different implementations. It is easy to see that the 
execution on GPUs is up to two orders of magnitude faster than on
CPUs. Moreover it is not surprising that the mobility versions of 
CPUs and GPUs exhibit lower performances.

The number of domains the system is decomposed into is not a multiple of the
number of compute units. Say the system is decomposed into $m$ domains and the
device provides $n$ compute units (see tables~\ref{tableGPU1}
and~\ref{tableGPU2}), with $m \mod n > 0$ and $m>n$. The device is completely
utilized for a fraction
\[
 f = \frac{m/n}{m/n+1}
\]
of time, during the remaining time only $m\mod n$ compute units are busy. An
increasing $f$ with $m$ leads to a sub-linear scaling of runtime with the system size.
In KMC we avoid this problem trough the aforementioned block-ahead of time updating 
method, which is not possible using dead border decomposition.

\begin{figure}[ht]
\begin{center}
\epsffile{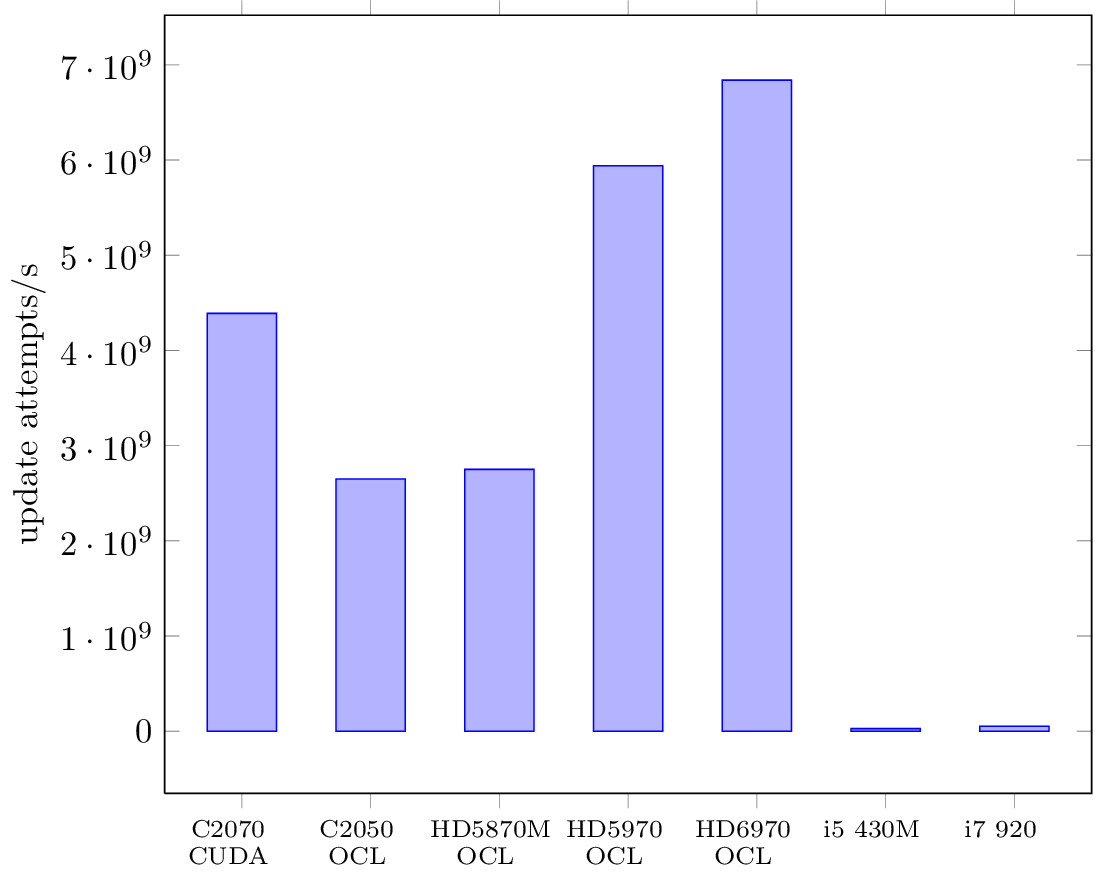}
\end{center}
\caption{Comparison of the computational speed in 
updates per second. For KPZ implementations. The values are in the order of
$\sim10^9$ for GPUs and $\sim10^7$ for CPUs.\label{fig:ups}
}
\end{figure}

In order to give a more illustrative idea on the performance differences 
between CPU and GPU codes, Figure~\ref{fig:ups} shows the number 
of updates per second on different architectures used. 
This value is a more practical one, because it really expresses 
the speed of a real physics application. 

\begin{figure}[ht]
\begin{center}
\epsffile{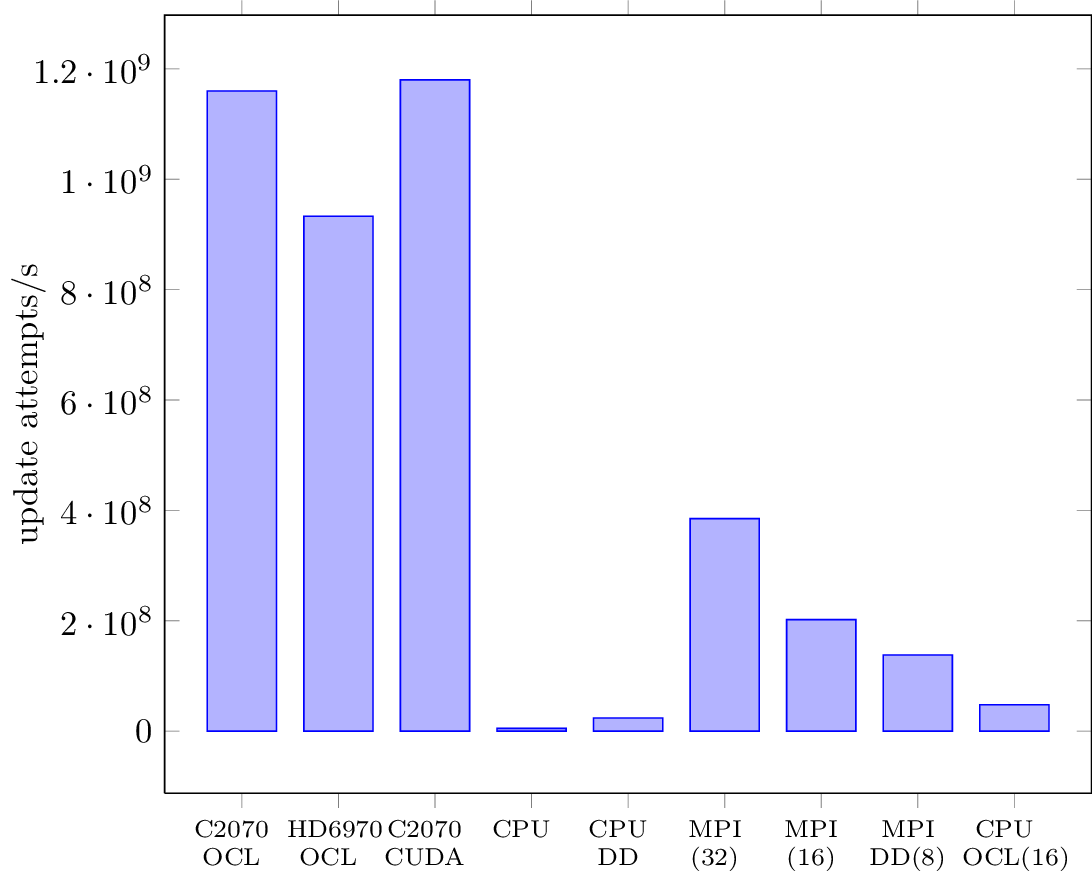}
\end{center}
\caption{Performance comparison of KMC implementations on GPUs 
and AMD Opteron CPUs using MPI or pure CPU code. The suffix 'DD' denotes L1
cache optimized CPU code. The numbers in parenthesis give the number of CPU
cores used for multi CPU implementations. The values are in the order of
$\sim10^9$ for GPUs and $\sim10^7$ for CPUs.\label{fig:mpi}}
\end{figure}

To benchmark our implementations of lattice KMC we simulated the quenching
process of a
system with an \texttt{fcc} lattice of $512^3/2$ sites. We started from a
homogeneous mixture with concentration of species $c=\num{.325}$
and effective temperature $\varepsilon=\num{1.5}$. Under these conditions
spinodal decomposition is observed. To provide some significance regarding real
world applications at least \SI{50}{kMCS} were performed. Since the workload
changes as phase separation and subsequent coarsening take place: The number of
successful update attempts decreases. Figure~\ref{fig:mpi} lists some of our
results.
The performance was normalized to the fastest single CPU implantation at hand.
The cache optimized version (CPU DD) was almost five times faster than the
classical CPU implementation on the AMD Opteron CPU used.

\begin{figure}
\begin{center}
\epsfxsize=120mm
\epsffile{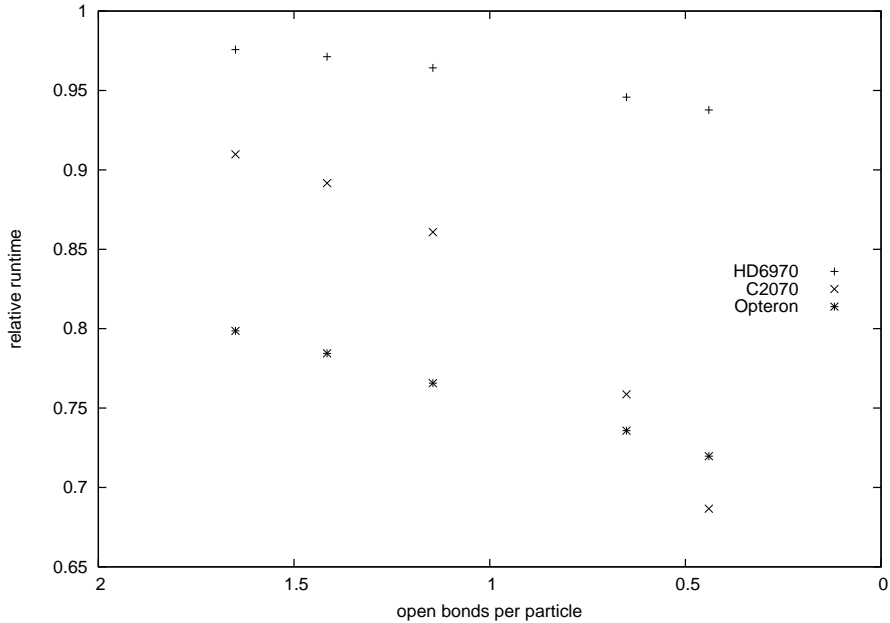}
\end{center}
 \caption{
 Run times of short runs (\SI{100}{MCS}) relative to run times using the initial
 state. The relative run times are plotted over the number of open bonds
 per particle, which is proportional to the internal energy of the system and a
 measure of relaxation. At the initial state for each specimen of B, around
 eight of it's twelve nearest neighbors are of species A (open bonds).
 \label{fig:rtLoad}}
\end{figure}

The straightforward OpenCL port of our CUDA implementation exhibits almost the
same performance on the same device, but it cannot utilize an HD6970 completely.
We also noted another difference between NVIDIA and AMD devices. As the
warp size on NVIDIA devices is smaller than on AMD devices, 
the C2070 can profit better from an increased rate
of  failing update attempts than the HD6970.
In fact, the HD6970 turned out to be slightly faster than a C2070 
for short runs, when almost all update attempts were successful.
Figure~\ref{fig:rtLoad} shows timing of short runs at late stages of the
evolution relative times measured at the initial state.

Our MPI implementation reaches an efficiency of $\approx\num{.5}$ for large
numbers of CPUs spread over multiple nodes. Because the system is only
decomposed in one direction no more that 32 cores can be used for a system of
the given size/aspect ratio. 
When all CPUs are located on a single node an efficiency of
$\approx\num{.73}$ is reached. In this case cache optimization was used.

Although it is possible to run OpenCL code on CPUs, it is not really efficient
with our application. Our tests show that OpenCL cannot keep its promise to enable
architecture independent parallel computing. Our code, designed for
NVIDIA GPUs, performs very badly on CPUs. Using 16 cores it delivers 
results only two times faster than a single core using cache optimization.

\section{Conclusions and outlook}

We have implemented and compared different GPU and CPU realizations using CUDA,
OpelCL and MPI for two applications: the $2+1$d KPZ surface growth and
phase separation of a binary, immiscible mixture in a 3d lattice KMC system.
We used dead border~\cite{GPU2cikk} and double
tiling domain decomposition, the latter turned out to be more efficient.
According to the benchmarks OpenCl performs almost as good
as CUDA on NVIDIA devices, but it could not fulfill it's promise on platform
independence. Using KPZ as an example we have shown, that adjusting 
the code to platform specifics has a large impact on performance 
and optimizing for one platform may result in performance penalties 
on the another. Our AMD-optimized OpenCl implementation of KPZ is 
considerably slower on NVIDIA GPUs than the straightforward port 
from CUDA. The same is true for running OpenCL code designed for GPUs 
on a CPU. Our MPI implementation of KMC outperforms our OpenCL 
implementation on CPUs.
Parallelizations of these applications by MPI on CPU clusters up to 
32 cores show low efficiency.

The cellular automaton approach in GPU simulations has been shown to be 
very efficient, because it allows bit coding as well 
as massively parallel computing.
This enabled us to simulate Ising type of models on unprecedentedly 
large scales both in space and time.
In particular we could study KPZ surface growth on sizes up to
$2^{17}\times 2^{17}$ lattice sites and a speedup $\sim 430$ was measured
on Fermi cards with respect to an I5-750 CPU core. For KMC
a speedup of $\sim70$ was found with respect to a Xeon E5530 core. 
We assume the lower speedup was due to the higher spatial
dimension rather than the complexity of KMC with respect to KPZ.
As we increase $d$, the number of lattice sites a work-item must
occupy grows. Since the local memory is limited, this decrases the number of 
work-items that can be run per block.
Furthermore, mapping of a higher dimensional system onto a linear array, the
locality of memory accesses decreases, increasing the possibility of
bank conflicts on GPUs.
For systems of equal volume, the latter should have no effect when running
on CPUs, as long as the system, respectively subsystem, fits into the cache.
The statistical analysis of the results has been presented elsewhere
\cite{GPU2cikk}.

This kind of surface mapping can be extended by other competing processes,
resulting in surface patterns. In particular, ripple and dot formation
has been studied in \cite{patscalcikk,iinmproc}. Implementation on GPUs
can lead to fast simulation of larger systems, which is essential in case
of the slow surface diffusion, quenched disorder~\cite{cpc11}, or by performing
aging studies~\cite{henkel2012}. 

\section{Acknowledgments}

We thank Nils Schmei{\ss}er for the useful discussions on parallelizing KMC
using MPI.
Support from the Hungarian research fund OTKA (Grant No. T77629),
OSIRIS FP7 and the bilateral German-Hungarian exchange program DAAD-M\"OB 
(Grant Nos. 50450744, P-M\"OB/854) is acknowledged.
The authors thank NVIDIA for supporting the project with high-performance
graphics cards within the framework of Professor Partnership.

\end{document}